\documentstyle[12pt]{article}
\newcommand{\la}{\label}

\newcommand{\M}{{\cal M}}

\newcommand{\non}{\nonumber}
\newcommand{\be}{\begin{equation}}
\newcommand{\ee}{\end{equation}}
\newcommand{\ba}{\begin{eqnarray}}
\newcommand{\ea}{\end{eqnarray}}
\newcommand{\bastar}{\begin{eqnarray*}}
\newcommand{\eastar}{\end{eqnarray*}}
\newcommand{\half}{{1 \over 2}}
\begin{document}
\begin{titlepage}

\begin{flushright}
UU-ITP-19/95 \\
HU-TFT-95-63 \\
hep-tt/9510198
\end{flushright}

\vskip 0.4truecm

\begin{center}
{ \bf \large \bf QUANTUM MECHANICAL ANOMALIES AND
\\ \vskip 0.2cm
   THE D{\bf \normalsize \bf E} WITT EFFECTIVE ACTION
\\
}
\end{center}

\vskip 0.8cm

\begin{center}
{\bf Topi K\"arki$^{*}$ } \\
\vskip 0.2cm
{\it Research Institute for Theoretical Physics \\
P.O. Box 9, FIN-00014 University of Helsinki, Finland}

\vskip 0.2cm
and \\
\vskip 0.2cm

{\bf Antti J. Niemi$^{\dagger\natural}$ } \\

\vskip 0.2cm

{\it Department of Theoretical Physics,
Uppsala University \\
P.O. Box 803, S-75108, Uppsala, Sweden $^{\ddagger}$ \\

{\rm and }\\

Department of Physics, University of British Columbia \\
6224 Acricultural Road,
Vancouver, B.C. V6T 1Z1, Canada } \\

\end{center}

\vskip 0.7cm
\rm
\noindent
We study the partition function of N=1
supersymmetric De Rham quantum mechanics
on a Riemannian manifold, with a nontrivial
chemical potential $\mu$ for the fermions.
General arguments suggest that when
$\mu \to \infty$ we should get the
partition function of
a free point particle. We investigate
this limit by exact evaluation of the fermionic
path integral. In even dimensions we find
the De Witt term with a definite numerical
factor. However, in odd dimensions our
result is pestered by a quantum mechanical
anomaly and the numerical factor
in the De Witt term remains ambiquous.

\vfill

\begin{flushleft}
\rule{5.1 in}{.007 in} \\
$^{\ddagger}$  \small permanent address \\ \vskip 0.2cm
$^{\dagger}$ \small Supported by G{\"o}ran Gustafsson
Foundation for Science and Medicine \\
\hskip 0.3cm and by NFR Grant F-AA/FU 06821-308
\\ \vskip 0.3cm
$^{*}$ {\small E-mail: \scriptsize
\bf TKARKI@FLTXA.HELSINKI.FI \\}
$^{\natural}$  {\small E-mail: \scriptsize
\bf ANTTI.NIEMI@TEORFYS.UU.SE}  \\
\end{flushleft}

\end{titlepage}

In a classic article \cite{dew} De Witt shows that
on a D-dimensional Riemannian  manifold $\M$ the path
integral action of a free point particle is
\be
S ~=~ \int \half g_{ab} {\dot x}^a {\dot x}^b
- \frac{\hbar^2}{\kappa} R
\la{deact}
\ee
The first term is the standard classical
contribution.
The second term is the scalar curvature
of $\M$ and it arises from quantum
corrections when the path
integral is derived from a second quantized
Hamiltonian. The numerical
parameter $\kappa$
reflects the inherent normal ordering
ambiguity that looms out when we pass
from the classical theory to its quantum
counterpart. Originally, De Witt \cite{dew} found
that $\kappa = 12$, but presently the
canonical value appears to be $\kappa = 8$
\cite{dew2}, \cite{nieuw} and it corresponds
to Weyl normal ordering of the second quantized
Hamiltonian. Furthermore,
arguments have been presented \cite{nieuw}
that (\ref{deact}) should be improved
to a noncovariant $ R \to R + \Gamma \Gamma$
where $\Gamma$ is the Christoffel symbol.
Obviously this inexactness
in path integral quantization
is quite provoking: The
propagator of a free point particle
should coincide
with the heat kernel of
the scalar Laplacian on $\M$, which is a
mathematically well-defined quantity \cite{bgv}.

In the present Letter we shall investigate
if a natural
value of $\kappa$ could be
substantiated from a first principles
computation. For this we
consider the N=1
supersymmetric  De Rham quantum mechanics
on $\M$ by adding a chemical
potential $\mu$ to the fermions.
A nontrivial $\mu$ breaks the supersymmetry
explicitly, and general arguments imply
that in the $\mu \to \infty$ limit
we obtain the scalar
Laplacian on $\M$. Hence
we expect that in this limit we
recover (\ref{deact})
but with a definite value for $\kappa$. We
should also be able to verify
if additional noncovariant
terms must be included in the path integral
action.

The N=1 De Rham quantum mechanics has been
studied extensively \cite{wit}, \cite{blau}.
The theory admits
an intrinsic geometric structure so
that powerful mathematical tools \cite{bgv}
become available:
The quantum mechanical Hilbert
space coincides with the exterior
algebra of $\M$,
and the supersymmetric Hamiltonian $H$ is the
(generalized) Laplacian
$\triangle$ that operates
on this exterior algebra,
\be
H ~=~ \triangle ~=~ d d^\star + d^\star d
\la{laplacian}
\ee
where $d$ is the nilpotent exterior derivative and
$d^*$ is its adjoint. Here ${}^\star$ denotes
Hodge duality and by construction it
leaves the Laplacian intact,
\be
H^\star ~=~ H
\la{1stlaw}
\ee
There are two species of fermions that
correspond mathematically to the
following two operations
on the exterior algebra:
The first operation we denote
by $i_{a}$, and it is
the contraction dual
to the basis of vector fields
$\partial_{a}$. The second
operation $\varepsilon^{a}$ is
the wedge multiplication by
the one forms  $dx^{a}$. These
two operations are Hodge duals to each other,
and in particular they satisfy
the standard fermionic
algebra
\[
i_{a}\varepsilon^{b} + \varepsilon^{b}
i_{a} = \delta_{a}^{b}
\]
Their commutator
\[
N=\half [ \varepsilon^{a} , i_{a}]
\]
defines a (normal-ordered) number operator
that counts the form-degree of a $n$-form
\[
N \omega ^{(n)} ~=~ n \omega^{(n)}
\]
and commutes with the Laplacian (\ref{laplacian})
\be
[ N , H ] ~=~ 0
\la{2ndlaw}
\ee
but under Hodge duality
\be
N^\star ~=~ D - N
\la{Nstar}
\ee

In the following we shall be interested in the
grand canonical partition function
\be
Z_G ~=~ STr \exp \{ -\beta (H+ \mu  N) \} ~=~
\sum_{n=0}^D (-z)^n Tr e^{-\beta
\triangle_n}
\la{grand}
\ee
where $z=e^{-\beta\mu}$ and $\triangle_n$ is the
restriction of the Laplacian on $n$-forms.
We note that for a generic $\beta$ and $\mu$
(\ref{Nstar})
implies that under Hodge duality
$\mu \stackrel{\star}{\to} - \mu$ and
\be
Z_G^\star ~=~ (-)^D \frac{1}{z^D} Z_G
\la{1stZ}
\ee
while (\ref{2ndlaw}) implies
that $Z_G$ remains
invariant under the shift symmetry
\be
\mu ~ \to ~ \mu \ + \  2 \pi i \frac{k}{\beta}
\la{2ndZ}
\ee
where $k$ is an integer.

For $\mu = 0$ standard supersymmetry
arguments imply that
(\ref{grand}) is
independent of $\beta$ and
tantamount to the Witten index of the N=1
De Rham theory. Indeed, since the number of
zero modes of $\triangle_n$ equals
the $n$th Betti
number $b_n$ of $\M$ {\it i.e.}
the dimension of the $n$th
cohomology class $H^n(\M,R)$,  this
means that for $\mu = 0$
(\ref{grand}) coincides with the Euler
characteristic of $\M$,
\be
Z_G (z=1) \ = \ Z \ = \
\sum_{n=0}^D(-1)^n Tr\;e^{-\beta
\triangle_n} \
\stackrel{\beta \to \infty}{\longrightarrow}
\ \sum_{n=0}^{D}(-1)^n b_n=\chi (M)
\la{euler}
\ee
We note that
Hodge duality implies that
when $D$ is odd (\ref{euler}) vanishes.

A nonvanishing $\mu$ breaks
the supersymmetry,
and now (\ref{grand})
depends nontrivially
on $\beta$ and $\mu$.
In the $\beta
\to \infty$ but $z$ fixed limit
it coincides with the
Poincar\'e polynomial
\be
Z_G ~=~ \sum_n (-z)^n b_n
\la{poincare}
\ee
while in the $\beta$ fixed but $\mu \to \infty$
($z\rightarrow 0$) limit we get
\be
Z_G ~ \stackrel{\mu \to \infty}{\longrightarrow} ~
Tr\ e^{-\beta \triangle_0}
\la{z0}
\ee
where $\triangle_0$ is the scalar Laplacian
on $\M$,
\be
\triangle_0 \ = \ -\frac{1}{
\sqrt{g}}\partial_{a}
\sqrt{g} g^{ab}
\partial_b
\la{lap0}
\ee
This means that the limit (\ref{z0})
coincides with the partition function
of a free point particle on $\M$.
(This is also our definition of the point
particle.)
In particular, the action in the
path integral representation of (\ref{z0})
should coincide with (\ref{deact}) for
a definite value of $\kappa$.

In the present Letter we shall
compute $\kappa$ by exact
evaluation of
the fermionic integrals
in the $\mu \to \infty$ limit of (\ref{grand}).
For this we introduce the
explicit representations
\ba
d \ &=& \ \varepsilon^{a}\partial_{a} \non \\
d^* \ &=& \ -g^{ab} i_{a}(\partial_{b}-
\Gamma^{c}_{bd}\epsilon^{d} i_{c}),
\la{ops}
\ea
where $\Gamma^a_{bc}$ is the Christoffel symbol,
\[
\Gamma^{a}_{bc}  = \frac{1}{2}g^{a
d}(\partial_{b} g_{cd} +
\partial_c g_{bd} - \partial_{d}g_{bc})
\]
For a Hamiltonian path integral we need a classical
canonical realization
of (\ref{ops}). We introduce
local coordinates $x^a$
on $\M$ and construct the cotangent
bundle by defining the Poisson brackets
\[
\{ p_{a} , x^{b} \} \ = \ -\delta^{b}_{a}
\]
We also identify
\bastar
\epsilon^{a} \ & \ \rightarrow \ & \ \psi^{a} \\
i_{a} \ & \ \rightarrow  \ & \ \overline{\psi}_{a},
\eastar
where $\psi$ and $\bar\psi$ are anticommuting
variables and impose the Poisson brackets
\bastar
\\
\{\overline{\psi}_{a},
\psi^{b}\} \ = \ -\delta^{b}_{a}
\eastar
We then have the following canonical realizations
\ba
d \ & \rightarrow &  \ Q \ = \ \psi^{a}p_{a} \non
\\
d^*  \ & \rightarrow & \ Q^* \ = \ - g^{ab}
\overline{\psi}_{a}(p_{b}-\Gamma ^{c}_{bd}
\psi^{d} \overline{\psi}_{c})
\non \\
N \ & \rightarrow & \ N \ = \ \psi^a \bar\psi_a
\ea
and for the classical Hamiltonian we find
\be
H = [d,d^*] \rightarrow =  \{ Q, Q^* \}=-g^{a
b}( p_{a}-\Gamma ^{c}_{ad} \psi^{d}
\overline{\psi}_{c} )(p_{b}-\Gamma ^{e}_{bf}
\psi^{f}
\overline{\psi}_{e} )-\frac{1}{2} R^{a
b} _{cd}
\psi^{c}
\overline{\psi}_{a} \psi^{d}
\overline{\psi}_{b}
\la{canham}
\ee
where
\[
R^{a}_{bcd} =
\partial_{c}\Gamma ^{a}_{bd}
- \partial_{d}\Gamma^{a}_{bc}
+\Gamma^{e}_{bd}\Gamma^{a}_{ec} -
\Gamma ^{e}_{bc}\Gamma ^{a}_{ed}
\]
is the Riemann tensor.
The canonical
action of the N=1 De Rham theory is then
\[
S=\int_{0}^{\beta} \dot{x}^{a}
p_{a} + \dot{\psi}^{a}
\overline{\psi}_{a} -H
\]
or by eliminating $p_a$
\be
S \ = \ \int \frac{1}{4}g_{ab}
\dot{x}^{a} \dot{x}^{b}+
\overline{\psi}_{a}
(\delta^{a}_{b}\partial_{t}
+\dot{x}^{c} \Gamma ^{a}_{cb})
\psi^{b}+\frac{1}{2} R^{ab}_{cd}
\overline{\psi}_{a}
\overline{\psi}_{b}
\psi^{c} \psi^{d}
\la{susyqm}
\ee
which is the standard action of the N=1
De Rham theory \cite{blau}. (The factor
of $\frac{1}{4}$ follows from
our normalization of $\triangle$.)
Notice that in contrast
to (\ref{deact}) here
we have {\it not} included the scalar curvature:
The corresponding (Hamiltonian)  path integral can
be evaluated
{\it exactly} by localization
methods \cite{palo}
and consistent with (\ref{euler})
it yields the Euler characteristic
{\it only} if the scalar curvature is absent.

In the present Letter
we are interested in the $\mu
\to \infty$ limit of the
grand canonical partition function
(\ref{grand}),
\be
Z_G ~=~ Str\; e^{-\beta (H +\mu N)} =
\int\limits_{PBC} [\sqrt{g} dx \, d\psi \,
d\overline{\psi}] e^{-S(\mu)}
\la{grand2}
\ee
where now
\be
S(\mu) \ = \ \int \frac{1}{4}g_{ab}
\dot{x}^{a} \dot{x}^{b} +
\overline{\psi}_{a}(\delta^{a}_{b}
\partial_t + \dot{x}^{c}\Gamma^{a}_{cb}
+ \delta^{a}_{b} \mu)\psi^{b} +
\frac{1}{2} R^{ab}_{cd}
\overline{\psi}_{a}\overline{\psi}_{b}
\psi^{c}\psi^{d}
\la{N}
\ee
and as in \cite{palo} we define the path integral
measure using mode expansions {\it w.r.t.}
some complete set of functions.
We observe that the Hodge duality (\ref{1stlaw}),
(\ref{1stZ}) is clearly visible in (\ref{N}), it
corresponds to the discrete transformation
\ba
\psi^a ~ & \leftrightarrow & ~ -
g^{ab}\bar\psi_b \non \\
\mu ~ & \rightarrow & ~ - \mu
\la{1stZ2}
\ea
On the other hand, the shift symmetry (\ref{2ndZ}) is
less transparent, it is a property of the path integral
(\ref{grand2}) and must be verified by an explicit
computation.

According to our general arguments the
$\mu \to \infty$ limit
of (\ref{grand2}) should coincide with the
path integral of a free point particle,
and we expect that in this limit
the effective action (\ref{deact}) emerges
with a definite value for $\kappa$.
Indeed, we shall find that when
$\mu \to \infty$
the fermion integrals in (\ref{grand2})
can be evaluated
exactly by
summing over diagrams that survive the
$\mu \to \infty$ limit.
For this we introduce anticommuting $c$-number sources
$\eta$ and $\bar\eta$, and consider
\bastar
Z_G[\eta, \, \overline{ \eta}] &
= & \int[d\overline{\psi}\,
d\psi]\exp \{ -\int
\overline{\psi}_a(\partial_t+\mu)
\psi^a + \overline{\psi}_a
\dot{x}^{c}\Gamma^a_{cb} \psi^b + \half
R^{ab}_{cd}\overline{\psi}_a
\overline{\psi}_b \psi^c \psi^d
\\
&+& \  \eta^a
\overline{\psi}_a + \overline{\eta}_a \psi^a  \} \non \\
&=& \int [d\overline{\psi}\, d\psi ] \exp\{ -\int
\overline{\psi}_a (\partial_t+\mu )\psi^a
+ {\cal L}_I(\overline{\psi}_a, \psi^b)
+ \eta^a \overline{\psi}_a + \overline{\eta}_a \psi^a \}
\eastar
\ba
&=& \exp \{-\int  {\cal L}_I (-\frac{\delta}{\delta
\eta^a}, -\frac{\delta}{\delta
\overline{\eta}_b })\}
\int[d\overline{\psi}\,d\psi ]\exp \{-\int
\overline{\psi}_a
(\partial_t+\mu )\psi^a + \eta^a \overline{\psi}_a +
\overline{\eta}_a \psi^a \} \non \\
&=& \exp\{ -\int  {\cal L}_I (- \frac{\delta}{\delta
\eta^a} , - \frac{\delta}{\delta
\overline{\eta_b}}) \} Z_0[\eta ,\overline{\eta}]
\la{pertexp}
\ea
This yields the following Feynman rules
for the vertices\\
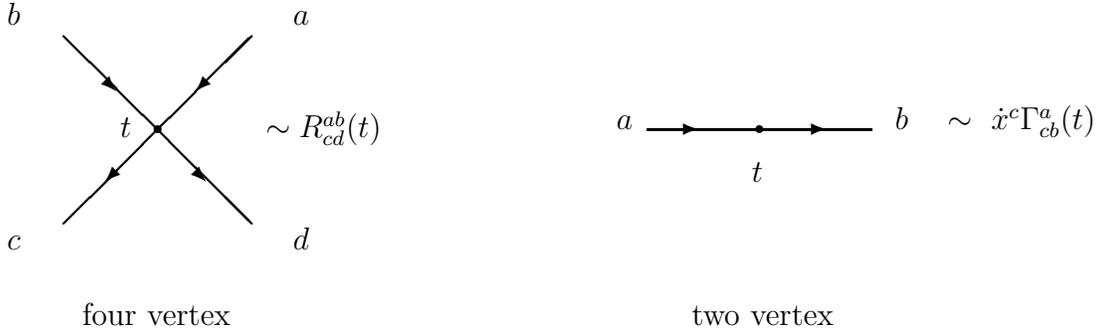
\begin{figure}[htb]
\setlength{\unitlength}{1mm}
\begin{picture}(155,55)
\thicklines
\put(12.5,20.5){\line(1,1){25}}
\put(12.5,45.5){\line(1,-1){25}}
\put(25,33){\circle*{1}}
\put(19.5,38.1){\vector(1,-1){0.0}}
\put(30.05,38.1){\vector(-1,-1){0.0}}
\put(18.0,26.0){\vector(-1,-1){0.0}}
\put(31.5,26.0){\vector(1,-1){0.0}}
\put(43,47){$a$}
\put(5,47){$b$}
\put(5,17){$c$}
\put(43,17){$d$}
\put(20,32){$t$}
\put(38,32){$ \ \sim R^{ab}_{cd}(t)$}
\put(15,7){$\rm four \ vertex$}
\put(105,33){\circle*{1}}
\put(90,33){\line(1,0){30}}
\put(97,33){\vector(1,0){0.0}}
\put(114,33){\vector(1,0){0.0}}
\put(86,33.08){$a$}
\put(123,33.08){$b$}
\put(104,26){$t$}
\put(130,33){$\sim \ \dot x^c \Gamma^a_{cb}(t)$}
\put(96,7){$\rm two \ vertex$}
\end{picture}
\end{figure}
\\
\noindent
and the propagator is determined by the Gaussian
$Z_0$ which evaluates to
\be
Z_0 \ =  \ C [Det_{\beta}(\partial_t+\mu)]^
D \ \exp \{ - \int
\overline{\eta}_a (t) <t|{ {\delta^a}_b \over
\partial_t+\mu } |t'> \eta^b (t') \}
\la{gaussian}
\ee
with $C$ a ($\mu$ independent)
normalization factor. For $t \not= t'$ the propagator is
\be
<t|{ \delta^a_b \over \partial_t+\mu }|t'> \ = \
\delta^a_b D(t'-t) \ = \ \delta^a_b
\sum\limits_{n=-\infty}^{\infty} { e^{2 \pi i n
\frac{t}{\beta} } \over
2 \pi i n  + \beta \mu} \ = \
\delta^a_b \left( \frac{1}{1-z }-
\theta (t-t') \right) e^{-\mu(t'-t)}
\la{mats}
\ee
and at $t = t'$ we can define
it self-consistently
using standard
identities that relate $D(0)$
to the determinant in (\ref{gaussian}),
\be
D(0) ~=~ \frac{1}{\beta} {\partial \over \partial \mu }
ln Det_\beta ( \partial_t + \mu )
\la{propalim}
\ee
In order to evaluate (\ref{propalim})
we need a regularization scheme.
However, as pointed out in \cite{elit} the determinant
is anomalous, there is no regularization that
preserves both the Hodge duality (\ref{1stZ2})
and the shift symmetry (\ref{2ndZ}).
Indeed, we find \cite{elit}
\be
Det_{\beta}(\partial_t+\mu) \ = \ {\rm sinh}
\half \beta \mu \cdot
e^{\phi \beta \mu} \ = \
\frac{1}{2} ( e^{\half \beta \mu} - e^{ - \half \beta
\mu} ) \cdot e^{\phi \beta \mu}
\la{det}
\ee
where $\phi$ parametrizes different
regularizations, and according to (\ref{propalim})
it also pesters $D(0)$,
\be
D(0) \ = \ \half
coth \half \beta \mu  \ + \
\phi ~ \ \stackrel{ \mu \to
\infty}{\longrightarrow}
\ ~ \half + \phi
\la{propalim2}
\ee

For $\phi = 0$ we have the Hodge duality
$\mu \to \ -\mu$ in (\ref{det})
but the shift symmetry (\ref{2ndZ})
is violated when $k$ is odd.
For $\phi = \pm\half$ we recover the shift
symmetry (\ref{2ndZ}) in (\ref{det})
but now the Hodge
duality is broken. For other values of
$\phi$ both symmetries are spoiled, consequently
the natural values of the parameter
are $\phi = 0$ and $\phi = \pm \half$.
Furthermore, since
the determinant (\ref{det}) appears
$D$ times in (\ref{gaussian})
we conclude that with $\phi = 0$ and $D$ even we
have anomaly cancellation and we recover
both Hodge duality and shift symmetry.
However, for $D$ odd the anomaly persists: When $\phi = 0$
(\ref{gaussian}) respects the Hodge duality but the
shift symmetry (\ref{2ndZ}) is broken for
odd values of $k$, while for
$\phi = \pm \frac{1}{2D}$ the shift symmetry
is recovered
but the Hodge duality is broken. Consequently
for even dimensional manifolds we select $\phi = 0$
and the anomaly
cancels in (\ref{gaussian}),
but for odd dimensional manifolds
the anomaly is unavoidable and
depending on $\phi$ it appears either in the Hodge
duality or in the shift symmetry, or in both.

We shall now proceed to evaluate (\ref{pertexp}) in
a diagrammatic expansion in the $\mu \to \infty$ limit.
For this we first expand
the path integral to second order. Diagrammatically,
\\
\begin{figure}[htb]
\setlength{\unitlength}{1mm}
\begin{picture}(155,50)
\thicklines
\put(5,37.3){$ Z_{\eta = \bar\eta = 0}  \ = \ (1-z)^D
\exp W  \ = \ (1-z)^D \exp \{ \ $}
\put(93,38){\circle{10}}
\put(88,38){\circle*{1}}
\put(97.8,37.1){\vector(0,-1){.0}}
\put(100.5,37.3){$+$}
\put(110.5,38){\circle{10}}
\put(120.5,38){\circle{10}}
\put(115.5,38){\circle*{1}}
\put(125.3,37.1){\vector(0,-1){.0}}
\put(105.5,38.8){\vector(0,1){.0}}
\put(128,37.3){$+$}
\put(138,38){\circle{10}}
\put(133,38){\circle*{1}}
\put(143,38){\circle*{1}}
\put(138.4,43){\vector(1,0){.0}}
\put(137.2,33.04){\vector(-1,0){.0}}
\put(20,17.3){$ + \ \ \ \ \ $}
\put(33,18){\circle{10}}
\put(43,18){\circle{10}}
\put(28.0,18.5){\vector(0,1){.0}}
\put(38,18){\circle*{1}}
\put(48,18){\circle*{1}}
\put(43.6,22.8){\vector(1,0){.0}}
\put(42.7,12.95){\vector(-1,0){.0}}
\put(52.5,17.3){$+$}
\put(64.5,18){\circle{10}}
\put(74.5,18){\circle{10}}
\put(84.5,18){\circle{10}}
\put(59.5,18.8){\vector(0,1){.0}}
\put(89.5,17.4){\vector(0,-1){.0}}
\put(69.5,18){\circle*{1}}
\put(79.5,18){\circle*{1}}
\put(74.7,23){\vector(1,0){.0}}
\put(74.2,13.05){\vector(-1,0){.0}}
\put(93.5,17.3){$+$}
\put(105.5,18){\circle{10}}
\put(105.5,18){\oval(9.9,3.5)}
\put(100.5,18){\circle*{1}}
\put(110.5,18){\circle*{1}}
\put(104.9,22.8){\vector(-1,0){.0}}
\put(106.3,13.07){\vector(1,0){.0}}
\put(104.9,19.65){\vector(-1,0){.0}}
\put(106.3,16.25){\vector(1,0){.0}}
\put(113,18){$\ + \ \ ... \ \}$}
\end{picture}
\end{figure}

\vfill\eject
\noindent
and explicitly,
\bastar
W &=& \int_0^{\beta}ds \ \left( \dot{x}^{a}\Gamma^b_{ab}(s)
D(0)+ R^{ab}_{ab} D(0)^2 \right)
\\
&-& \frac{1}{2} \int_0^{\beta} ds\,ds' \
\dot{x}^{a} \Gamma^b_{ac} (s') \dot{x}^{d}
\Gamma^{c}_{cb} (s) D(s'-s) D(s-s')
\\
&-& 4 \int_0^{\beta}ds  dt \ \dot{x}^{a}\Gamma^b_{ac}(s)
\half R^{cd}_{bd}(t) D(t-s)  D(0)  D(s-t)
\\
&-& 8\int_0^{\beta}dt\,
dt' \half R^{ab}_{cb}(t) \half
R^{cd}_{ad} ( t') D(t-t') D(0)^2
D(t'-t)
\\
&+& 2 \int_0^{\beta} dt\,dt' \half
R^{ab}_{cd} (t) \half R^{cd}_{ab} (t') D(t-t')^2 D(t'-t)^2
\end{eqnarray*}

We argue that in the $\mu\rightarrow\infty$
limit only the following two diagrams survive\\
\begin{figure}[htb]
\setlength{\unitlength}{1mm}
\begin{picture}(155,30)
\thicklines
\put(5,13.3){$W(\mu \to\infty) \ = $}
\put(41,14){\circle{10}}
\put(36,14){\circle*{1}}
\put(45.8,13.1){\vector(0,-1){.0}}
\put(48.5,13.3){$\ +$}
\put(62,14){\circle{10}}
\put(72,14){\circle{10}}
\put(67,14){\circle*{1}}
\put(76.8,13.1){\vector(0,-1){.0}}
\put(57.1,14.8){\vector(0,1){.0}}
\put(82,13.3){$ = \ \int\limits_0^\beta
\dot x^a \Gamma^b_{ab} (t)
D(0) \ + \ R^{ab}_{ab} (t) D(0)^2$}
\end{picture}
\end{figure}
\\
\noindent
For this we consider an arbitrary
connected diagram. Its
propagator structure is
\be
D(\tau_1-\tau_2)D(\tau_2-\tau_3)\cdots
D(\tau_I-\tau_1)
\la{dig1}
\ee
where the $\tau_i$ are not necessarily
all distinct. Here $I$ is the
number of internal lines which we have ordered so
that they constitute a solution to Euler's
bridge problem, we walk around the diagram in
the direction of the arrows in such a way that we
pass each internal line exactly once.
This is possible because we only have
vertices with an even ({\it i.e.} either 2 or 4)
number of lines. We recall that in a
given diagram the number $V_n$ of
$n$-vertices is connected to the number of
internal lines $I$ by the topological
relation
\[
I ~=~ V_2 + 2 V_4
\]
and the number of loops is
\[
L ~=~ I - V_2 - V_4 + 1
\]
Using the explicit form (\ref{mats})
we then find that the
exponential factors $e^{-\mu t}$ in (\ref{dig1})
all cancel, and we are left with
\[
(\frac{1}{1-z}-\theta(\tau_2-\tau_1))\cdots
(\frac{1}{1-z}-\theta(\tau_1-\tau_{I-T})) \cdot
D(0)^T
\]
where $T$ is the number of tadpoles.
As $z\to 0$ this
reduces to
\[
\stackrel{z \to 0}{\longrightarrow}~
\theta(\tau_1-\tau_2)\cdots
\theta(\tau_1-\tau_{I-T}))
\cdot D(0)^T
\]
which is nonvanishing only if
the number of tadpoles coincides
with the number of internal line. Hence
$V_2 + V_4 \leq 1$ as we asserted, and
we have established that
\[
\int[d\overline{\psi}\,d\psi]\exp \{-\int
\overline{\psi}_a
(\nabla _t+\mu)\psi^a+ \half R^{ab}_{cd}
\overline{\psi}_a \overline{\psi}_b
\psi^c\psi^d\}
\]
\[
\stackrel{\mu \to \infty}{\longrightarrow} \
\exp \{ \int dt(\hbar
\dot{x}^{a}\Gamma^b_{ab} (t) D(0) +
\hbar^2 R^{ab}_{ab} (t) D(0)^2)\}
\]
where we have re-introduced $\hbar$ to
count the number of loops.
Since $\dot{x}^{a}\Gamma^b_{ab}
=\frac{1}{2}\partial_t \ln \det
g_{ab}$ is a divergence the first term does not
contribute to (\ref{z0}),
and we conclude that the
path integral representation of
(\ref{z0}) is
\[
Tr \, e^{-\beta \triangle_0} =
\int[\sqrt{g}dx] \exp\{
-\int \frac{1}{2}g_{ab}\dot{x}^a\dot{x}^b
- \frac{\hbar^2}{2} (\half + \phi)^2 R^{ab}_{ab}\}
\]
where we have redefined $\beta \to \half \beta$.
This coincides with the De Witt action (\ref{deact})
with $\kappa = \frac{1}{8} (1 + 2 \phi)^2 $.
In particular, the result depends nontrivially
on $\phi$ in (\ref{det}).

We have already concluded that
for even dimensional
manifolds the natural value
is $\phi = 0$ which ensures that
the anomaly in the
determinant (\ref{det})
cancels. For the (Euclidean)
path integral action this yields
\[
S ~=~ \int \frac{1}{2}g_{ab}\dot{x}^a\dot{x}^b
- \frac{\hbar^2}{8}  R^{ab}_{ab}
\]
Hence on even dimensional manifolds we obtain
the De Witt action (\ref{deact}) with
$\kappa = 8$, fully consistent with the $\beta \to 0$
result in \cite{dew2}.

For odd dimensional manifolds the anomaly in
(\ref{det}) persists and we have three natural
values of $\phi$, either the Hodge
dual $\phi = 0$
or the shift symmetric $\phi = \pm \frac{1}{2D}$.
Thus we find that the
De Witt term appears with numerical factors
$\kappa = 8$ and $\kappa = \frac{1}{8}\frac{D^2}{
(D \pm 1)^2 }$
respectively. Obviously the analogy with
$D$ even suggests that we should
select the Hodge dual $\kappa = 8$
also in odd dimensions, but this would
be at the expence
of the shift symmetry.

Finally we point out, that unlike the
construction in \cite{nieuw} our approach is
manifestly covariant. In particular, we do
not find any evidence for terms such as
$\Gamma \Gamma$ in (\ref{deact}).
However, since we have not
attempted to evaluate the remaining bosonic
path integrals in (\ref{grand2}),
we can not exclude the possibility
that noncovariant counterterms are necessary
to properly define the bosonic path
integral measure.

\vskip 0.5cm
In conclusion, we have investigated the N=1
supersymmetric De Rham theory with
a supersymmetry breaking chemical
potential for the fermions. General arguments
suggest that in the $\mu \to \infty$
limit we should
recover the De Witt effective path integral
action with a definite
numerical factor. In even dimensions we indeed
find De Witt's result with the numerical
factor $\kappa = 8$. But in odd dimensions
the $\mu \to \infty$ limit is plagued by a quantum
mechanical anomaly and $\kappa$
admits {\it three} natural values, including the
Hodge symmetric $\kappa = 8$.

\vskip 1.0cm
T.K\"arki thanks Mikko Laine, Mauri
Miettinen and Kaupo Palo
for discussions.

\end{document}